\documentclass[12pt]{article} 
\usepackage{psfig}

\newcommand{\plotan}[3]{ {\begin{figure}[h]
\begin{minipage}[t]{9cm}\makebox[0cm]{}\\
\psfig{file=#1.eps,width=9cm} \end{minipage} \hfill
\parbox[t]{4.8cm}{\makebox[0cm]{}\\ \caption{\label{#3} \sf #2}
}
\end{figure}
}}
\newcommand{\e}{\mbox{e}}

\begin{document}

\title{Modeling sublimation by computer simulation:\\ morphology
	 dependent effective energies} 
\author{S. Schinzer and W. Kinzel\\
Universit\"at W\"urzburg, Institut f\"ur Theoretische Physik\\
Am Hubland D-97074 W\"urzburg, Germany}

\maketitle

\begin{abstract}
Solid-On-Solid (SOS) computer simulations are employed to investigate
the sublimation of surfaces. We distinguish three sublimation regimes:
layer-by-layer sublimation, free step flow and hindered step flow. The
sublimation regime is selected by the morphology i.e. the terrace
width. To each regime corresponds another effective
energy. We propose a systematic way to derive microscopic parameters
from effective energies and apply this microscopical analysis to the
layer-by-layer and the free step flow regime.  We adopt analytical
calculations from Pimpinelli and Villain and apply them to our model.

{\em keywords:} Computer simulations; Models of surface kinetics;
Evaporation and Sublimation; Growth; Surface Diffusion; Surface
structure, morphology, roughness, and topography; Cadmium telluride 
\end{abstract}

\section{Introduction}
The sublimation of semiconductor materials such as cadmium telluride
or silicium has been of considerable interest in the last few
years. The motivation of these studies has been mainly to understand
the kinetics on these surfaces. This would be helpful e.g. to gain
better control in the fabrication of nanostructure devices. In this
paper we want to describe computer simulations of sublimation.  We use
the conceptually simple SOS model which has been extensively used to
model Molecular Beam Epitaxy (MBE) \cite{sv95,lp96}.  As far as we
know these are the first simulations of sublimation of this kind.

Theoretical studies of sublimation were based on the 1+1 dimensional
Burton-Cabrera-Frank model. In this model the evolution of the surface
is described by the motion of steps. In 2+1 dimensions it can be
interpreted as a surface of parallel, mono-atomic and straight steps.
Pagonabarraga et. al. examined the validity of the continuum
approximation in the step flow regime \cite{pve94}. Pimpinelli and
Villain discussed the creation of surface vacancies and found a
condition for the occurrence of ``Lochkeime'' (poly-vacancies)
\cite{pv94}. In the following we will refer to calculations in their
introductory book of crystal growth \cite{vp95}. In particular they
found that in step flow sublimation the velocity of steps is
influenced by the terrace width.

Experimental studies of CdTe(001) concentrated on the extraction of
effective energies. Several authors used Reflection High Energy
Electron Diffraction (RHEED) to study sublimation
\cite{blw95,tdb94,umm91}.  Other investigations were based on the use
of a quadrupole mass spectrometer (QMS) \cite{nts97b,jfh90} (and
references therein). Recently Neureiter et al applied Spot Profile
Analysis of Low Energy Electron Diffraction (SPA-LEED) to
CdTe(001)\cite{nts97a,nts97b}. The oscillation period $\tau$ in RHEED
or SPA-LEED intensity can be described by an arrhenius-law $1/\tau =
\nu \; \e^{-E/kT}$.  For CdTe(001) an activation energy of $1.9eV$ -
derived from the diffraction data - is commonly accepted . Using QMS a
value of $1.55 eV$ for the mass desorption rate has been reported.
We will show that this discrepancy can be explained by the existence
of different morphology dependent sublimation regimes. Neureiter
et. al. show that such a picture of the CdTe(001) surface is
consistent with SPA-LEED and QMS measurements\cite{nts97b}.

It would be desirable to obtain microscopic parameters of a computer
model from ab-initio calculations. Calculations are now available for
GaAs(001) \cite{ks96ip,bs97}, but for II-VI semiconductors no such
calculations are available yet. Thous computer simulations of MBE or
sublimation can help to get insight into these
parameters. E.g. \v{S}milauer and Vvedensky were able to find a set of
parameters for GaAs(001).  They compared the step density of their
model to RHEED intensities during MBE growth and subsequent recovery
\cite{sv93}. These investigations showed that it is possible to describe
such a complicated surface (reconstruction, 2 types of atoms ...) by
means of the simplest SOS model. Naturally the microscopic parameters
found are still {\em effective} parameters.  We expect that these
values should be related to the true microscopic interactions
describing the slowest processes.

There is a certain advantage in using sublimation data compared to MBE
to find microscopic parameters of real materials. First of all the
process is determined uniquely by microscopic processes, whereas in
MBE the evolution of the morphology is dominated by the external flux
of atoms. Secondly, sublimation has been recently studied by SPA-LEED
\cite{nts97a,nts97b}. To
a great extent this method can be analyzed in kinematic approximation
\cite{wol95,hen85}, whereas RHEED-profiles have to be calculated using
multiple-scattering theory \cite{km97}. Also the mass desorption
during sublimation can be easily compared with computer simulations.

The organization of this paper is as follows. In section two we will
introduce the SOS model and the algorithm used in the simulations. In
section three we will investigate sublimation similar to experimental
investigations. We will identify three sublimation regimes in
dependence of the terrace width. In section four we will connect
effective activation energies (which are measurable in real
experiments) to the underlying microscopic parameters.  On the one
hand we will be able to compare these relations with theoretical
predictions of Pimpinelli and Villain (section five). On the other
hand this leads to a systematic approach for determining microscopic
energies of real materials which will be illustrated in section six.
A discussion and an outlook will follow in section seven.

\section{Solid-On-Solid model}
One basic assumption in this model is the neglection of
overhangs. Thus no bulk vacancy formation can be investigated. The
second assumption is a fixed lattice structure. Both lead to the
crucial simplification for the simulation that the surface can be
stored in one simple data-array of heights.

\plotan{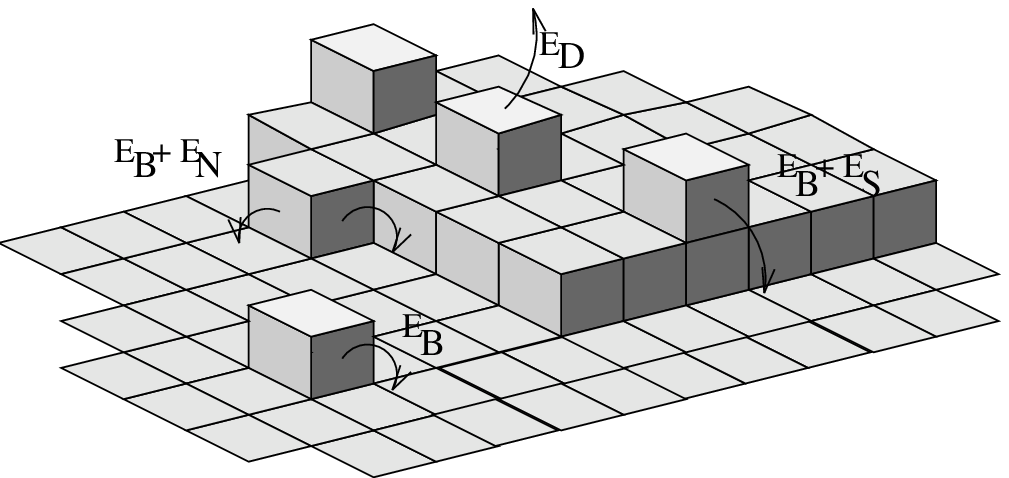}{The activation energy in the jump rate of atoms depends
on the local arrangement. The diffusion barrier is $E_B$, for
desorption an energy of $E_D$ is necessary. Next in-plane neighbours
have a binding energy of $E_N$. A jump over a step edge is hindered
by the Ehrlich-Schwoebel barrier $E_S$.}{SOS}

We have used a simple version of this model: a simple cubic lattice
with only one species of atoms. Since there is only one lattice
constant all lengths will be given in units of this lattice constant.
Figure \ref{SOS} shows some selected processes and their corresponding
activation energies.

Theories of activated processes lead to an arrhenius form of the jump
frequency $\Gamma = \nu_0 \: \e^{-E/kT}$ \cite{gle41}. The prefactor
in general depends on the form of the potential and the
temperature. Nevertheless the exponential factor gives the dominant
temperature dependence. Thus we assume a constant prefactor
$\nu_0$. The activation energy depends on the microscopic mechanism
involved. We assume that diffusion has an activation energy of $E =
E_B + n\cdot E_N$ where $n$ is the number of next in-plane
neighbours. Thus free atoms in this model have the diffusion constant
$D = \nu_0 \: \e^{-E_B/kT}$. At step edges an Ehrlich-Schwoebel
barrier $E_S$ is considered. In addition to ``conventional''
SOS-models of MBE we take into account the desorption similar to the
diffusion with an jump frequency $\Gamma_{desorption} = \nu_0 \:
\e^{-(E_D+n\cdot E_N)/kT}$. We assume the same prefactor for
desorption as for diffusion which is certainly not valid in real
materials.

Especially when simulating sublimation it is absolutely important to
include all possible events because even events with very low rates
can become very important. E.g. the creation of a surface vacancy has
a rate which is several orders of magnitudes lower then the jump
frequency of free adatoms. Nevertheless this process is the most
important one to initiate layer-by-layer sublimation.

The Maksym-algorithm \cite{mak88} is able to handle such big
differences in jump rates. In this algorithm all possible events are
considered simultaneously and are stored in one array of rates. The
event with a high rate will be chosen with a corresponding high
probability. Maksym described how to find in an efficient way the
event selected by a random number in the range $(0,\sum rates)$. We
improved the simulation of the (time consuming) diffusion of free
adatoms by handling them separately.

In this paper we will concentrate on two sets of parameters: $\nu_0 =
10^{12} s^{-1}, \:\: E_B=0.9eV, \:\: E_D = 1.1eV, \:\: E_N = 0.25eV,
\:\mbox{ and }\: E_S = 0.1eV$ (set 1). A second set $\nu_0 = 10^{12}
s^{-1}, \:\: E_B=0.5eV, \:\: E_D = 0.7eV, \:\: E_N = 0.35eV, \: \mbox{
and }\: E_S = 0.1eV$ (set 2) will be compared.

\section{Crossover from layer-by-layer to step flow sublimation}
Since computer simulations are restricted to rather small systems (we
use system sizes up to $800 \times 800$ lattice constants, typically
$512 \times 512$), one is forced to induce equidistant steps by fixing
a height difference between two opposite boundaries. We have
investigated the effect of such an induced step train on the mass
desorption.

There are two limiting cases. First the pure layer-by-layer
sublimation. In this case the poly-vacancies are much smaller then
typical terrace sizes. Changing the terrace sizes in this regime
should not influence the total amount of layers desorbed after a fixed
time. Furthermore the number of layers desorbed should be equal to the
number of periods observed in LEED or RHEED.

In the second extreme the steps are so close together that there is no
place to create poly-vacancies and there will be only desorption from
steps. In this case the number of desorbed monolayers $\Delta h$ after
the time $\Delta t$ will be equal to
\[ \frac{\Delta h}{\Delta t} = \frac{v_{step}}{L_{terrace}}\]
because $L_{terrace}/v_{step}$ is just the time to evaporate one
monolayer.

\plotan{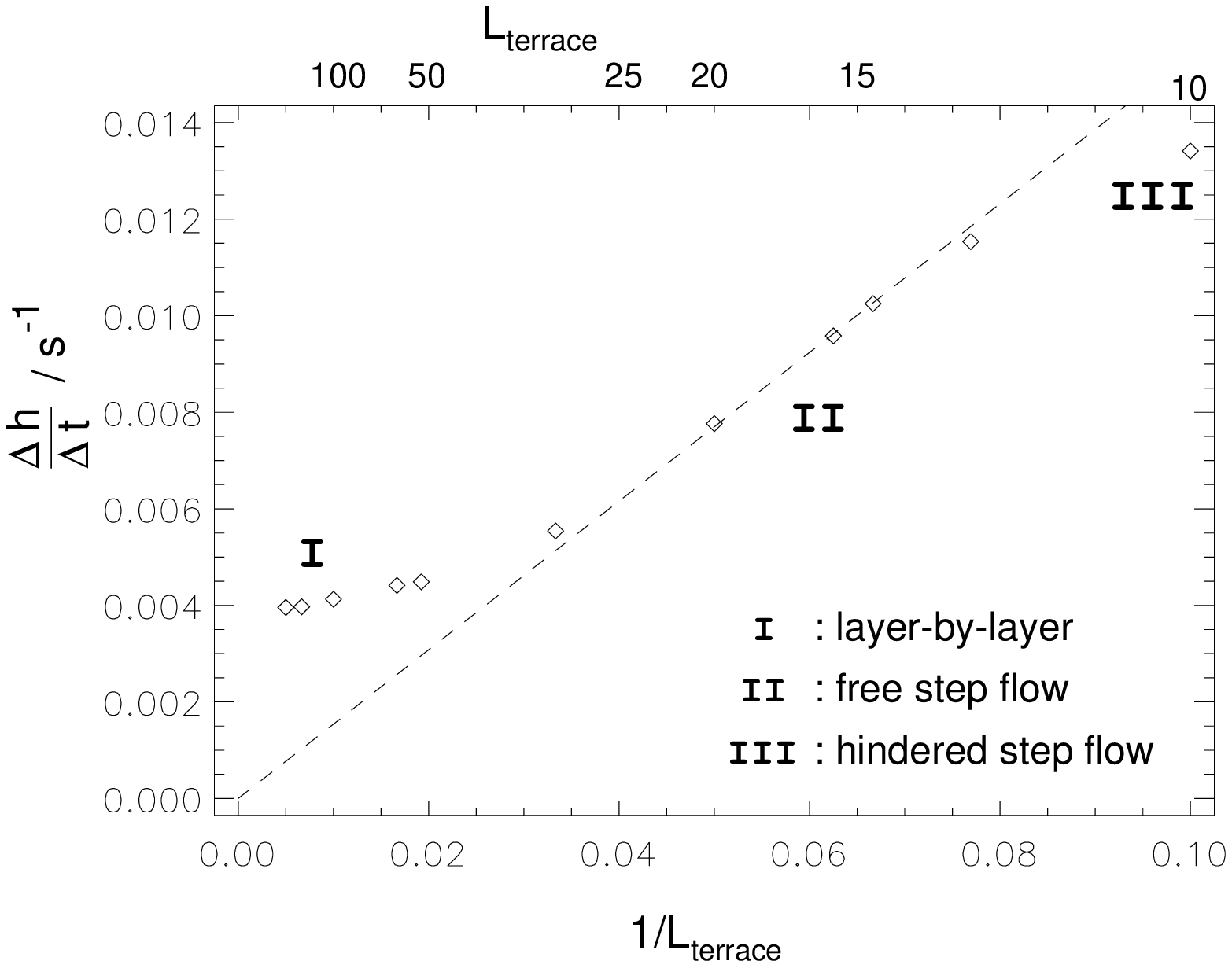}{Height loss per time over $1 / L_{terrace}$.  Each
data point represents a simulation of parameter set (1) at $600 K$.
For all simulations $\Delta t$ was chosen to be $800 s$. The dashed
line corresponds to $v_{step}=0.154 s^{-1}$.}{dh}

Figure \ref{dh} clearly show the existence of these two
regimes. However at very small terrace sizes a third regime can be
identified (we have not plotted another data point at $1/L_{terrace} =
0.2$ which clearly establishes region III). This behaviour was already
calculated by Pimpinelli and Villain using the modified
Burton-Cabrera-Frank (BCF) model\cite{vp95}. In this framework but
without a Schwoebel-effect (however the estimation is valid in
general) $v_{step} \propto \tanh (L_{terrace}/2 \sqrt{D \tau_D})$,
 where $D$ is the diffusion constant of free atoms and $\tau_D$ the
life-time of an particle until desorption occurs. If the terrace width
is greater then the diffusion length $\sqrt{D \tau_D}$ one gets a constant
velocity. But if one lowers $L_{terrace}$ the particles will cross the
terraces and will be reincorporated at the step edges. In this case
$v_{step}$ will be proportional to $L_{terrace}$ and $\Delta h/ \Delta
t$ will become constant again. That is why we
want to distinguish between hindered and free step flow. Setting
the argument of the tanh to one and identifying
$D=\nu_0 \; \e^{-E_B/kT} \mbox{ and } 1/\tau_D = \nu_0 \;
\e^{-E_D/kT}$ one can estimate the crossover length between regime
{\tt II} and {\tt III} to be $\ell_\times
\approx 14$ in good correspondence to figure \ref{dh}.

Likewise Pimpinelli and Villain calculated the crossover between the
free step flow and the layer-by-layer regime. In the approximation
described in section five we obtain for parameter set (1) a crossover
length of about 60, which is once again in good correspondence to our
simulations. However we will show in section five that this agreement
of theory and simulation is merely by chance.

In addition we have investigated the intensity in antiphase of the
(0,0)-spot using the kinematic approximation.  We found that the
oscillation period was the same either by calculating the step density
or the width of the height distribution (in the case of singular
surfaces).  The evaluation of the LEED-profiles is complicated by the
fact that due to the small system sizes the distribution of terrace
sizes is limited. This leads to a pronounced twin-peak structure in
k-space at $k_{peak} = \pm \frac{2 \pi}{L_{terrace}}$. We have
mimicked the effect of a broad distribution of terrace sizes by
convolution of the obtained profile with an Gaussian of $0.5\%$ of the
Brioullin Zone. In correspondence to the conclusions drawn from the
mass desorption diagram (fig. \ref{dh}) one oscillation period of $232
s \pm 8$ down to a terrace size of $L_{terrace} \approx 100$ is
observed.  At $L_{terrace} = 64$ an oscillation could be identified in
the twin peak structure. However the period could only be estimated to
be roughly $\tau \approx 265s \pm 20s$. With $L_{terrace} = 32$ no such
oscillations occurred.

We have shown that in regime {\tt I} (layer-by-layer sublimation) the
observed oscillations are independent of the terrace size. Decreasing
the mean terrace width one reaches the free step flow regime (regime
{\tt II}) which is characterized by a constant step
velocity. Furthermore oscillations do not exist in this
regime. However at small terrace sizes (small compared to the diffusion
length) one reaches the hindered step flow regime {\tt III} which is
again characterized by a mass desorption rate independent of the terrace
width.

\plotan{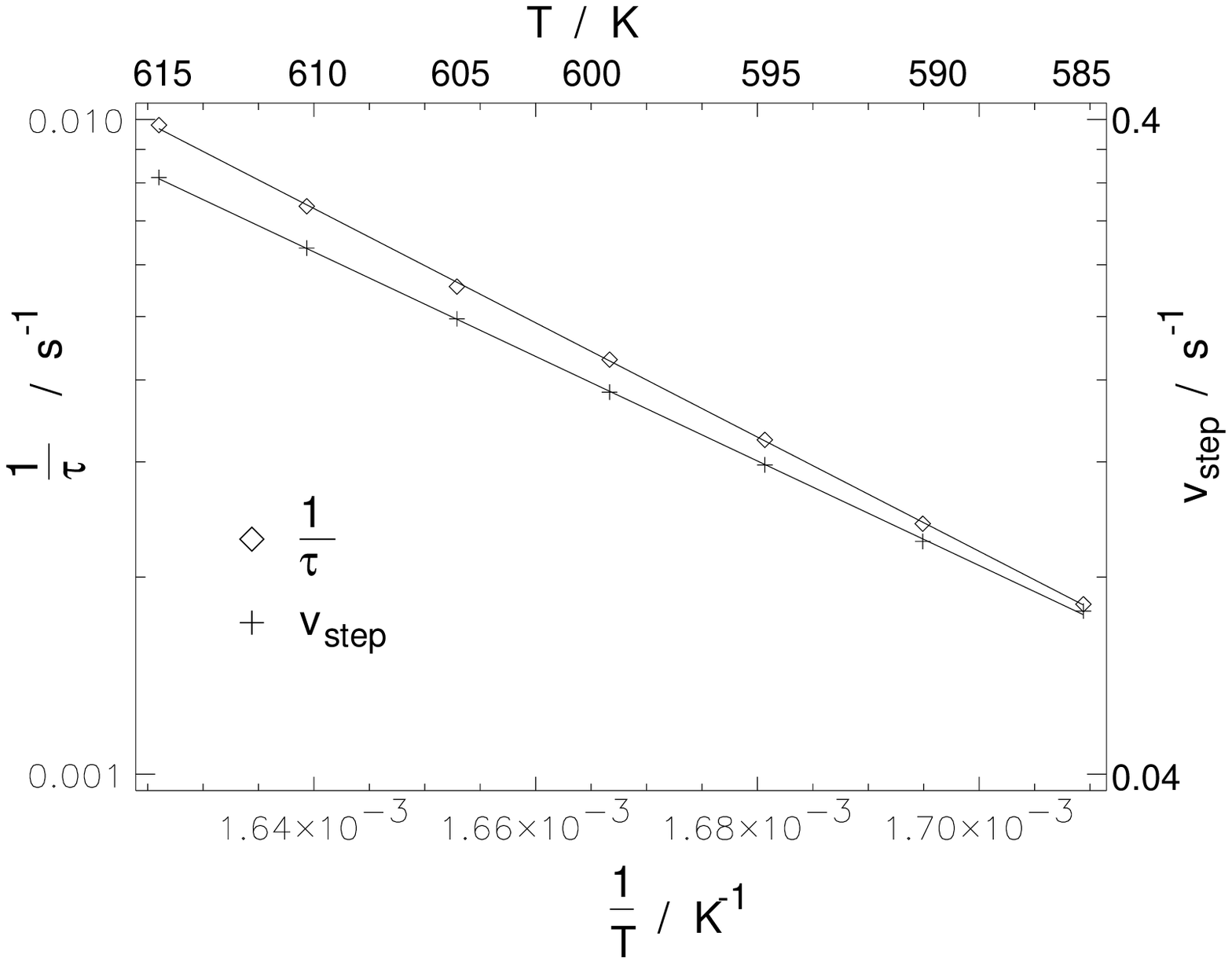}{Comparison of the activation energies of the
layer-by-layer and the free step flow sublimation of parameter set
(1). In the layer-by-layer regime ($L_{terrace} = 256$) the
oscillation period is characterized by the effective activation energy
$E_{I}=1.73 eV$. At $L_{terrace}=16$ the step velocity yields
$E_{II}=1.60 eV$.}{compare}

In each regime one could expect to define effective energies.  In
regime {\tt I} the oscillation period is a well defined variable.  In
regime {\tt II} -the free step flow regime- the step velocity and at
small terrace sizes (regime {\tt III}) the mass desorption rate can be
analyzed. We concentrate on regime {\tt I} and {\tt II} but we will
mention the theoretical result (including the
Ehrlich-Schwoebel-barrier) in section 5.3 for regime {\tt
III}. Looking at the temperature dependence of $1/\tau$ and $v_{step}$
(figure \ref{compare}) one can indeed identify arrhenius laws with
different activation energies.

\section{Interpretation of effective energies: microscopical analysis}
We will now try to establish a link between the microscopic parameters
of the model to the effective energies. The effective energies are
derived from the variation of the temperature. In the simulations
however it is possible to vary the microscopic parameters as well.

We will vary the energies of set (1) by fixing three energies and
changing the fourth. In detail we have set the parameters to
$E_B/eV$=\{0.8, 0.85, 0.9, 0.95\}, $E_D/eV$=\{1.00, 1.05, 1.10,
1.15\}, $E_N/eV$=\{0.23, 0.24, 0.25, 0.26\} and $E_S/eV$=\{0.05, 0.10,
0.15, 0.20\}.  Assuming a linear dependence of the effective energy on
the parameters $E_{\mbox{\small \tt regime}} = \beta E_B + \delta E_D + \nu E_N +
\sigma E_S$ we want to extract the importance of the different
microscopic parameters.

\subsection{Layer-by-layer sublimation (regime {\tt I})}
Our simulations at different temperatures (s. figure \ref{compare} )
show an arrhenius behaviour with $\nu^T_{I} = 1.48\;10^{12}\;s^{-1}$
and an effective activation energy of $E^{T}_{I} = 1.73 \;eV$. The
upper index ``T'' indicates that these are values obtained by
simulations at different temperatures.

The investigation of the effect of the microscopic parameters yields
$\nu^{l.a.}_{I} = 6.3\;10^{11}\;s^{-1}$ and
\[E^{l.a.}_{I} = 0.61\cdot E_B + 0.35\cdot E_D + 2.85\cdot E_N +
0.44\cdot E_S .\]

In our case this leads to an effective activation energy close to the
``conventional'' effective energy
\[ E^{l.a.}_{I} = 1.69 eV \approx E^{T}_{I} = 1.73eV. \] 
Note that $E^{l.a.}_{I}$ has been deduced from simulations at a fixed
temperature $T=600 K$.  At first glance it is surprising that the
diffusion parameter $E_B$ has a larger influence on the effective
energy than the desorption energy $E_D$. This can be explained by the
typical life of an evaporating particle. Firstly the atoms become
freely diffusing atoms (this needs an activation energy of typically
$E_B+3 \cdot E_N$. Thereafter the atoms diffuse on the terrace until
they evaporate. The Ehrlich-Schwoebel barrier plays an essential role
in the creation of surface vacancies which explains the influence of
$E_S$.

\plotan{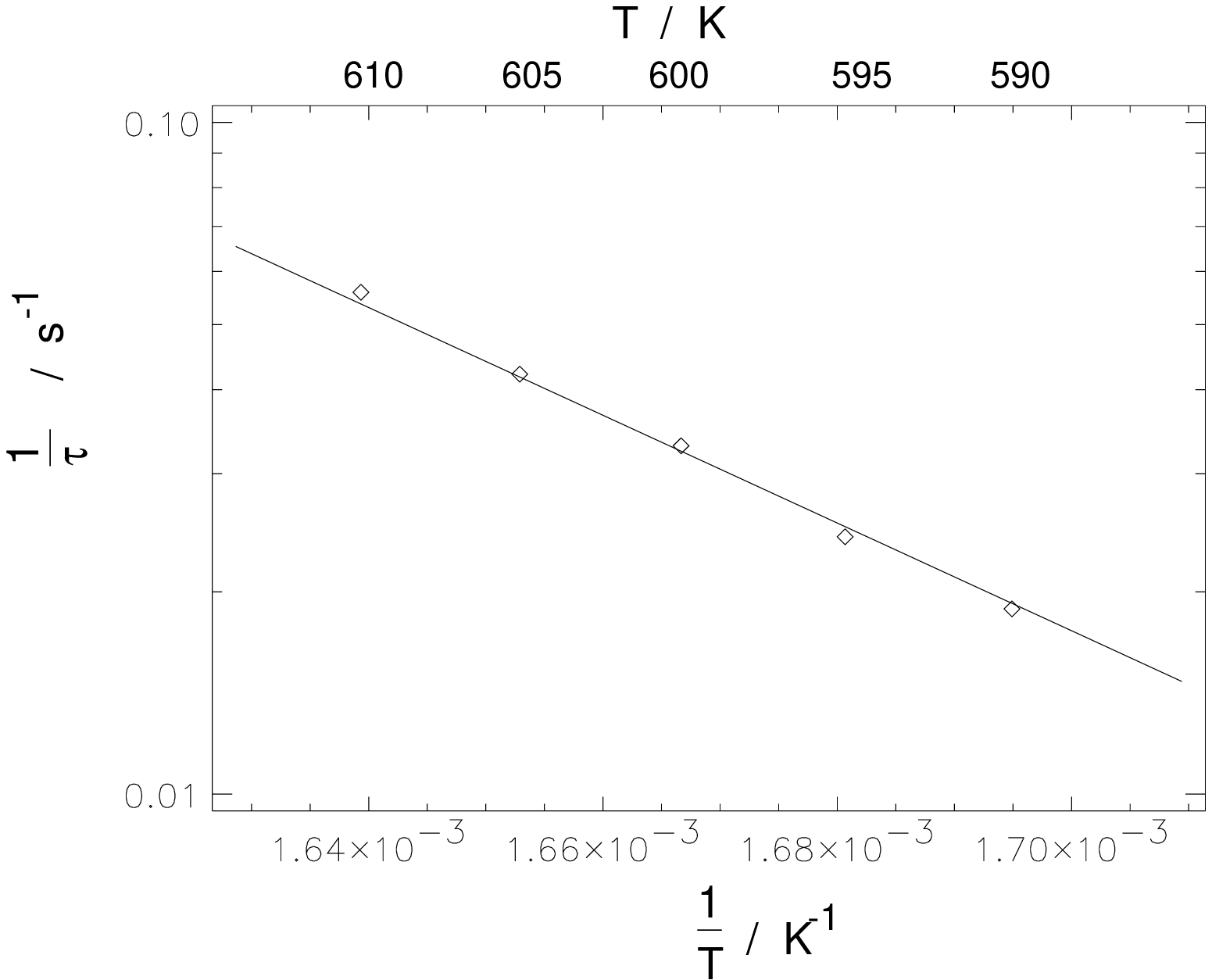}{Temperature dependent sublimation rate in the
layer-by-layer regime of set (2). However the predicted curve was
multiplied by a factor 1.2 to show that the predicted energy is very
good.}{model2_lbl}

\begin{table}[h]
\begin{tabular}{c|c|c|c||c|c}
$E_B/eV$&$E_D/eV$&$E_N/eV$&$E_S/eV$&$\tau_{period}/s$&$\tau_{predicted}/s$
\\ \hline 0.82 & 1.20 & 0.16 & 0.2 & 9.2 & 3.2 \\ 0.50 & 0.70 & 0.35 &
0.1 & 28 & 37 \\ 0.75 & 0.80 & 0.30 & 0.1 & 89 & 88 \\ 0.42 & 0.42 &
0.42 & 0.1 & 122 & 103 \\ 1.15 & 1.30 & 0.20 & 0.1 & 1290& 1181\\
\end{tabular}
\caption{\label{pre-sim-loch} Predicted and measured
Oscillation periods. The simulations were carried out at $T = 600 K$,
the vibration frequency was set to $\nu_0 = 12^{12} s^{-1}$ }
\end{table}

Other simulations show, that this approximation of first order is
valid over a greater range in parameter space. The predicted periods
of these models are compared with measured ones in Table
\ref{pre-sim-loch}.  Figure \ref{model2_lbl} shows that the activation
energy predicted by our linear approximation is still very good for
the set (2) although the prefactor no longer holds.

\subsection{Free step flow (regime {\tt II})}
In the following investigations we used a surface with a mean terrace
size of 16 lattice constants. Looking at Figure \ref{dh} one can see
that one is in the free step flow regime.

By evaluating the temperature dependence in this extreme limit we find
$\nu_{II}^T = 3.9 \cdot 10^{12} \; s^{-1}$ and $E_{II}^T = 1.60 eV$.

As in the case of layer-by-layer sublimation we connected this
effective relation to the microscopic parameters. Of course we cannot
be sure to stay in the free step flow regime. We checked that $\Delta
h / \Delta t$ is at least twice as high as in the pure layer-by-layer
regime. However with high $E_D$ or low $E_B$ we reached the hindered
step flow regime due to the fact that the atoms stayed to short a time
on the terrace.

Our results can be fitted by
\[ E_{II}^{l.a.} =  
  0.70 \cdot E_B + 0.32 \cdot E_D + 2.4 \cdot E_N + 0.21\cdot E_S \]
In this case the two different effective energies are identical
$E^{l.a}_{II}=E^{T}_{II}=1.60eV$. Likewise the prefactor of $4.3 \cdot
10^{12} \; s^{-1}$ is close to the previous one.

Comparing this relation with the equivalent relation for the
layer-by-layer regime one observe that the contribution of $E_N$ and
$E_S$ have significantly changed. This seems plausible because the
creation of a surface vacancy compared to the creation of a kink is
hindered by the energy $E_N+E_S$.

\plotan{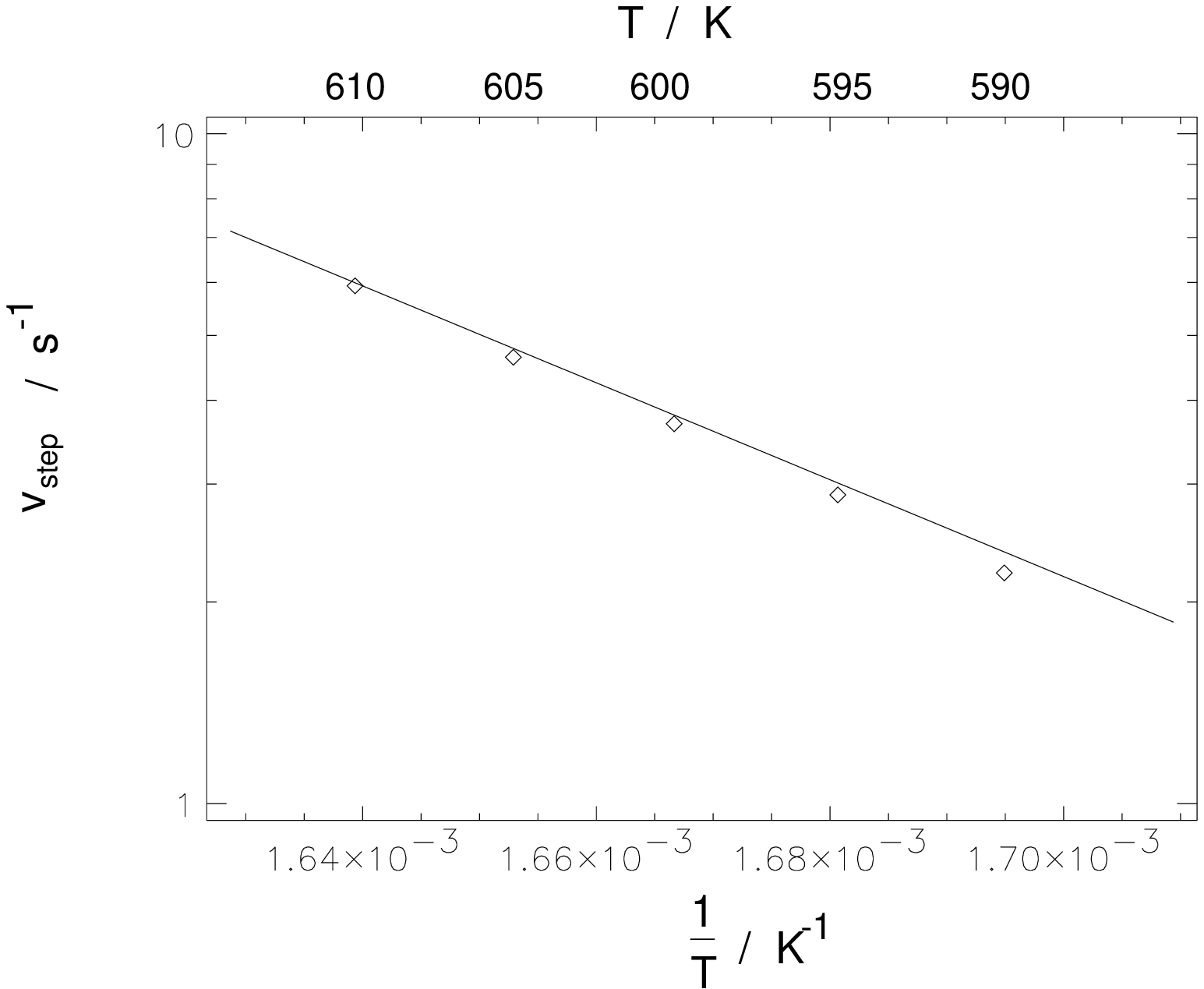}{Predictions and measurements of $v_{step}$ for
parameter set (2). No corrections were necessary!}{model2_fsf}

A comparison between our prediction for the set (2) and measurements
are shown in figure \ref{model2_fsf}. As one can see our linear
approximation is very good in this case.

\section{Comparison with theoretical calculations}
Now we are able to compare the theoretical predictions of Villain and
Pimpinelli with our results. We will interpret variables in their
results in terms of arrhenius activated jump rates. 

\subsection{Free step flow (regime {\tt II})}
In the free step flow regime they
obtained the following expression for $v_{step}$ on a surface with
terraces of equal size $\ell$ (ch. 6.4 in \cite{vp95})
\[ v_{step} = \rho_0 \kappa D
\frac{4-\left( 2 - \kappa D
\left( \frac{1}{D''}+ \frac{1}{D'} \right) \right) \e^{-\kappa \ell}
-\left( 2 + \kappa D \left( \frac{1}{D''}+ \frac{1}{D'} \right)
\right) \e^{\kappa \ell}}
{\left(1+\kappa \frac{D}{D'}\right)
\left(1+ \kappa \frac{D}{D''}\right) \e^{\kappa \ell} - \left(1-\kappa
\frac{D}{D'}\right) \left(1- \kappa \frac{D}{D''}\right) \e^{-\kappa
\ell}}. \] 
 $\rho_0$ can be identified with the ratio between the detachment
rate from a step and the diffusion jump rate ($\rho_0 = \e^{-(n\cdot
E_N)/kT}$, where $n\cdot E_N$ is a typical binding-energy to next
in-plane neighbours at a step). The diffusion parameters at step
edges are related with the diffusion constant $D$. Diffusion downward a
step $D'=\e^{-E_S/kT}D$. Diffusion towards a step from the lower terrace
is not changed in our model ($D''=D=\nu_0 \cdot \e^{-E_B/kT}$) . The
inverse of the diffusion length is given by
$\kappa=1/\sqrt{D\tau_D}=\e^{-(E_D-E_B)/2kT}$. Taking the limit
for $\kappa \ell \gg 1$ (free step flow) and neglecting 
$\e^{-(E_D-E_B)/kT}$ versus 1 we obtain the simpler relation
\[ v_{step} \approx \nu_0 \e^{-\frac{E_B/2 + E_D/2 + n\cdot E_N}{kT}}
	\cdot \left( 1 + \frac{1}{1+\e^{-(E_D/2-E_B/2-E_S)/kT}}
\right). \] 
This shows that the assumption of an arrhenius behaviour is not valid
in this sublimation regime. However with parameter set (1)
the correction factor has no temperature dependence . Because
$E_D/2-E_B/2 = E_S$ we obtain 
 $v_{step} \approx 3/2 \: \nu_0 \e^{-(E_B/2 + E_D/2 + n\cdot E_N)/kT}$.
In the microscopical analysis we measure an effective
importance of $E_D$, $E_B$ and $E_S$ due the correction factor. The
influence of $E_N$ is unchanged. Indeed if we use the theoretical
result with $n \cdot E_n = 2.4 \cdot E_N$ (as the microscopical
analysis suggests) we obtain an effective energy of $1.6 eV$ in
correspondence to our measurement.

\subsection{Crossover to layer-by-layer sublimation}
Given the two equations for the period of oscillations and the step
velocity we can calculate in our linear approximation approach the
dependence of $L_\times$ from the temperature and the microscopic
parameters. From $ 1/\tau = v_{step}/L_\times$ we obtain
\[ L_\times = \frac{\nu_{fsf}}{\nu_{lbl}} \e^{-(E_{fsf}-E_{lbl})/kT} 
	\approx 6.8 \e^{(0.03 E_D - 0.09 E_B + 0.45 E_N + 0.23
E_S)/kT}. \] We already mentioned the theoretical relation for the
crossover terrace width. Pimpinelli and Villain estimated the
crossover to be at
\[ \ell_\times^2 \ln \ell_\times \approx 4 \pi
\left(\frac{\gamma}{kT}\right)^2 \tau_D D \left(1+ \frac{\Lambda
\sigma_0}{D \rho_0}\right). \] $\sigma_0$ is the equilibrium
concentration and $\Lambda$ the diffusion constant of the surface
vacancies. $\gamma$ is the free energy per step length. Neglecting
$\Lambda$ and setting $\gamma = E_N$ (which is a good approximation if
the step is nearly free of kinks) we can relate this expression to our
model.  If we simplify their relation in the vicinity of $\ell_\times$
at $T=600K$ we obtain approximately
\[ \ell_\times \approx \sqrt{\pi} \left( \frac{E_N}{kT} \right) \cdot
e^{(E_D-E_B)/kT}. \] which is at the same order of magnitude as our
result but has a very different temperature dependence. In the
theoretical result only desorption and diffusion energies contribute
to the exponential temperature dependence. Whereas our simulations
indicate that these terms cancel out and only $E_N$ and $E_S$
contribute exponentially.

\subsection{Hindered step flow (regime {\tt III})}
We have not examined in detail the hindered step flow
regime by computer simulations. Nevertheless we want to mention the
theoretical result for 
this regime. In this regime $\kappa \ell \ll 1$ and therefor one can use
the Taylor series of the exponential functions $\e^{\pm \kappa \ell}
\approx 1 \pm \kappa \ell$. In this case the step velocity is described
by an arrhenius behaviour.
\[ v_{step} \approx \ell \nu_0 \e^{-(E_D+n \cdot E_N)/kT} \]
This means that the number of monolayers desorbed is independent of
the terrace width in this regime. 

\section{Systematic derivation of microscopic energies}
Our connection between effective and microscopic energies will help to
develop a set of parameters describing e.g. the CdTe(001)
surface. Taking the effective energies for CdTe(001) as $E_{I} = 1.9
eV$ and $E_{II}=1.55 eV$\footnote{It is possible that the mass
desorption energy is due to regime {\tt III}. We just want to
demonstrate how the microscopical analysis can be used.} we can then
reduce the four dimensional
energy-parameter space to a two dimensional one. E. g. the
Schwoebel-barrier can be expressed in terms of the diffusion and
desorption barriers: $E_S = 0.31 eV + 1.16 E_B + 0.16 E_D$. In figure
\ref{EnEbEd} the $E_N(E_B,E_D)$-plane is shown. But because there are
further physical restrictions (namely that $E_B>0$, $E_N>0$ and
$E_D>E_B$) only a triangle is left from the infinite plane. 
One should however be careful. In the dark region (where
$E_D<E_B+E_S$) surface vacancies will be created via direct desorption
rather than the creation of an adatom-vacancy pair. In this
region our microscopic analysis is certainly incorrect.
It should be emphasized that our relation of effective to
microscopic energies is only valid in the vicinity of parameter set
(1). However one can take it as a first approximation. Then one can
systematically improve the result by iterating the microscopic
analysis.

\plotan{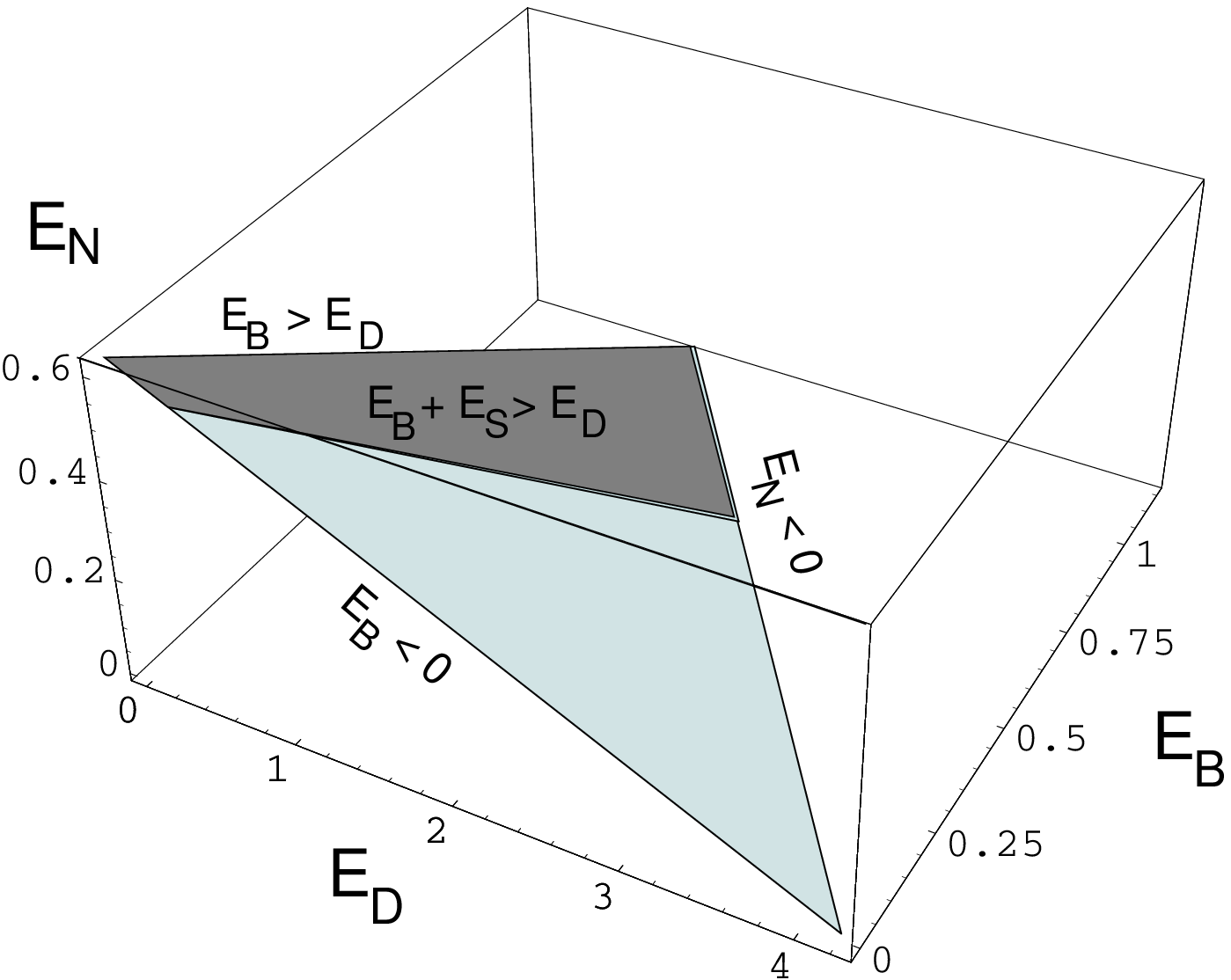}{Proposed parameter plane $E_N(E_B,E_D)$ for CdTe(001). Due
to physical constraints a triangle is cut out. In the dark region
our microscopical analysis is not valid. There the process of
the generation of surface vacancies is changed.}{EnEbEd}

\section{Conclusion}
In this paper we have reported the first simulations of sublimation in
the framework of the SOS model. We have been able to identify three
regimes of sublimation by looking at the mass desorption:
layer-by-layer sublimation, free and hindered step flow.  The
existence of these regimes is in correspondence to predictions for the
BCF-model from Pimpinelli and Villain.

We have related the corresponding effective activation energies to the
underlying microscopic parameters. Although our relation is just a
linear approximation in the vicinity of the parameter set (1) the
application to other sets of parameters is quite successful. From
these relations we were able to compare the step velocity and the
cross-over to layer-by-layer with analytical results from Pimpinelli
and Villain. Our result for the step velocity is similar to the
analytical result. However our approximation for the crossover length
is very different from the analytical one.

Our simulations show that the effective desorption energy is to a
great extent influenced by the diffusion of adatoms. Thus energies
measured in real experiments cannot be understood as desorption
energies. Furthermore the energies measured are highly influenced by
the morphology of the surface. We have only compared singular and
stepped surfaces but it is now obvious that e.g. in MBE one has to
expect other desorption rates than measured by Sublimation.

In this work we have concentrated on the sublimation period and mass
desorption. Morphological features as typical surface vacancy
distances will be the work of further investigations. This way we hope
to get a more detailed insight in the physics of sublimating surfaces.

\section*{Acknowledgement}
We would like to thank Michael Biehl, Herbert Neureiter and Moritz
Sokolowski for fruitful discussions. This work has been supported by
the Deutsche Forschungsgemeinschaft through SFB 410.

\bibliography{Literatur}
\bibliographystyle{unsrt}

\end{document}